\documentclass[aps,prb,twocolumn,groupedaddress,floatfix,showpacs]{revtex4}

\usepackage{graphicx,latexsym}

\begin{document}


\title{The Stability of Phase Coexistence in Atomic Clusters}


\author{S. C. Hendy}
\affiliation{MacDiarmid Institute for Advanced Materials
and Nanotechnology, Industrial Research Ltd, Lower Hutt, New Zealand}



\date{\today}

\begin{abstract}
Microcanonical critical droplet theory and molecular dynamics simulations are used to examine
static coexistence between solid and liquid phases in nanoscale lead clusters. It is shown that the
theory predicts the existence of a metastable coexisting state above a critical cluster radius
$R_1$, with this state becoming stable for clusters of radius $R > R_2$. Molecular dynamics
simulations of lead clusters confirm the existence of stable coexisting states in 1427-atom
and 2057-atom clusters but find no stable coexisting state in a 931-atom cluster.
\end{abstract}

\pacs{61.46.+w,64.70.Dv,68.08.-p}

\maketitle

\section{Introduction}
The coexistence of solid and liquid phases in atomic clusters has been of experimental
\cite{Pochon04,Lee04} and theoretical \cite{Honeycutt87, Reiss88, Wales94} interest for some 
time. In large clusters, as in bulk materials, static coexistence is expected
theoretically\cite{Nielsen94} and is readily observed in both experiment\cite{Pochon04,Lee04} 
and simulation\cite{Cleveland94}.
However in sufficiently small clusters, simulations find that coexistence is dynamic i.e.
clusters fluctuate in time between fully solid and fully liquid states \cite{Honeycutt87}. Small
clusters avoid static coexistence because of the prohibitive cost of forming an interface. This
causes an S-bend in the microcanonical caloric curve \cite{Lynden-Bell} and the
corresponding negative heat capacities which been observed in small sodium clusters \cite{Schmidt01}.

However, it is currently not known at what size static coexistence gives way to dynamic coexistence.
The time-scale on which dynamic coexistence occurs will be governed by the height of the energy barrier separating the solid and
liquid states of the cluster \cite{Reiss88}. However, if a coexisting state exists, then this
coexisting state should also be dynamically accessible. This is not seen in at least some small
clusters \cite{Honeycutt87,Cleveland99}. Evidently, in sufficiently small clusters coexistence is
unstable. At what sizes do states of static coexistence occur?

To determine when phase coexistence will be stable, one needs to know the relative cost of forming an
interface between phases in a given cluster. A common approach to studying coexistence in clusters
is to apply critical droplet theory in the capillarity approximation (i.e. where cluster material
parameters, especially surface energies, are approximated by their bulk values).
For instance Reiss et al \cite{Reiss88} used the
capillarity approximation to develop a critical droplet theory of melting in the canonical ensemble,
with the (spherically-symmetric) solid fraction of a cluster acting as an order parameter. Nielsen et
al \cite{Nielsen94} developed a similar critical droplet theory in the microcanonical ensemble. They
demonstrated that a stable static coexisting state can exist in this theory near the melting point,
and validated this with constant-energy simulations of a large copper cluster. Cleveland et
al \cite{Cleveland94} developed an aspherical model which dealt with static coexistence where the
liquid only partially wets the solid. Thus, in principle, critical droplet theory can be used to
discuss coexistence in a variety of circumstances.

To study coexistence we will work in the microcanonical ensemble as constant temperature ensembles 
tend to suppress the inherent inhomogeneity in coexisting systems\cite{Gross01}. Constant energy 
conditions are realised in isolated systems with poor thermal coupling to their environment, such as
clusters in an inert gas atmosphere. Consequently we will use microcanonical critical droplet (MCD)
theory \cite{Nielsen94} to examine the stability of static solid-liquid coexistence in small metal
clusters. We will then use constant-energy molecular dynamics simulations to test the predictions of 
the theory.

\section{Microcanonical Critical Droplet Theory}

Here we use microcanonical critical droplet (MCD) theory \cite{Nielsen94} to examine the stability of
static coexistence. The spherical coexisting solid-liquid cluster of radius $R$, is considered to
consist of a solid core, of radius $R_s$, and a liquid outer shell. The entropy (per atom), $S_m$, of
a coexisting solid-liquid cluster of solid fraction $\eta=(R_s/R)^3$ at energy (per atom) $e$ is
approximated by
\begin{equation}
\label{entropy}
S_m(e)=\eta S_s(e_s)+(1-\eta) S_l (e_l)
\end{equation}
where $S_s(e_s)=S_s^c + c \log\left({e_s-e_s^c \over c T_c} +1\right)$ and
$S_l(e_l)=S_l^c + c \log\left({e_l-e_l^c \over c T_c} +1\right)$. Here $T_c$ is the bulk melting
temperature, $c$ is the heat capacity and quantities with a superscript $c$ indicate values at
the bulk melting point. The solid energy, $e_s$, and liquid energy, $e_l$, are related to the total
energy, $e$, for a cluster of radius $R$ as follows:
\begin{eqnarray}
\label{energy}
e & = &\eta e_s + (1-\eta)e_l  \\
&+& {3 \over \rho R} \left(\gamma_{lv} + (\gamma_{sl}
+\Delta \gamma \exp\left(-2(1-\eta^{1/3})R/\xi\right))\eta^{2/3} \right) \nonumber
\end{eqnarray}
where $\Delta \gamma = \gamma_{sv}-\gamma_{sl}-\gamma_{lv}$ ($\gamma_{sv}$ is the solid-vapour
interfacial energy density, $\gamma_{sl}$ is the solid-liquid interfacial energy density and $\gamma_{lv}$ is the
liquid-vapour interfacial energy density), $\rho$ is the bulk material density and $\xi$ is a
correlation length which characterizes the short-range interaction between the solid-liquid and liquid-vapour
interface \cite{Nielsen94}. The inclusion of this final term in equation (\ref{energy}) ensures
that $e \rightarrow e_s + 3 \gamma_{sv}/ \rho R$ as $\eta \rightarrow 1$.

We now show that MCD theory predicts that there is a critical cluster radius below which coexistence
becomes unstable. For a given total energy per atom $e$, we can regard $e_l$ as a function of
$e_s$ and $\eta$: $e_l = e_l (e_s,\eta)$. We may then extremise the entropy (given by (\ref{entropy}))
of the mixed cluster with respect to $e_s$ and $\eta$ at fixed $e$. It is then straightforward
show that the entropy is extremised when the following conditions hold: 1) $T_l = T_s$, which is the
normal condition for thermal equilibrium, and 2) $\eta$ must satisfy the following equation
\cite{Nielsen94}:
\begin{widetext}
\begin{equation}
\label{quartic}
\eta^{4/3}-\frac{3 \eta}{\rho R L} \left(\gamma_{sl} +
\Delta \gamma e^{-2(1-\eta^{1/3})R/\xi}\right) +
(e-e_l^c-\frac{3\gamma_{lv}}{\rho R}
+\frac{2 c T_c \Delta \gamma}{\rho \xi L} e^{-2(1-\eta^{1/3})R/\xi} )\frac{\eta^{1/3}}{L}
+ \frac{2 c T_c}{\rho R L^2} \left(\gamma_{sl}+\Delta \gamma e^{-2(1-\eta^{1/3})R/\xi} \right)= 0.
\end{equation}
\end{widetext}
where L = $e_l^c - e_s^c$ is the latent heat of fusion per atom. For sufficiently negative values
of $e$, there are two positive solutions to (\ref{quartic})
$\eta_1(e) > \eta_2(e) > 0$. There is always a range of energies where $0< \eta_2(e) < 1$, as
when $e \rightarrow -\infty$, $\eta_2(e) \rightarrow 0$. In fact $\eta_2(e)$ represents a
local minima in the entropy, corresponding to a barrier separating the solid and liquid phases.
Similarly, for values of $e$ where $\eta_1(e)$ exists and is less than one, $\eta_1(e)$ is a local
maxima in the entropy, corresponding to a coexisting state (note that $\frac{d e}{d \eta}
|_{\eta_1} < 0$ and $\frac{d e}{d \eta} |_{\eta_2} > 0$).

A range of energies where $0 < \eta_1(e) < 1$ exists only when $R > R_1$, where $R_1$ is given by:
\begin{equation}
\label{R1}
R_1 = \frac{\xi}{2} \left(-A+\sqrt{A^2+B}\right),
\end{equation}
where $A=\frac{3 \rho L^2}{4 c T_c}
\left(\xi - 2 \frac{\Delta \gamma}{\rho L} \left(1-\frac{2 c T_c}{3L} \right)\right)$ and
$B=2\left(1+\frac{3L}{c T_c}\right)(\gamma_{sv}-\gamma_{lv})$. That is if $R > R_1$, there is
a range of energies where a locally stable coexisting state exists. At such an energy e,
the coexisting state, with solid fraction $\eta_1(e)$, is separated from the liquid state
by an energy barrier at $\eta_2(e)$.

However, the coexisting state with solid fraction $\eta_1(e)$ is not necessarily globally
stable. If the coexisting state exists at an energy $e$, then $S_m(e) > S_s(e)$. However, there
exist a range of energies where $S_m(e) > S_l(e)$ only if $R > R_2$, where
\begin{equation}
\label{R2}
R_2 = \frac{\gamma_{sv}-\gamma_{lv}}{\rho L}
\left(\frac{3L-2 c T_c\left(1-e^{-L/c T_c}\right)}
{L- c T_c \left(1-e^{-L/c T_c}\right) \left(1-\frac{2 \Delta \gamma}{\rho \xi L} \right)} \right).
\end{equation}
Thus for clusters with $R_1 < R < R_2$, at energies where the coexisting state exists, the state
is always metastable  ($S_m(e) < S_l(e)$). For clusters with $R < R_1$, no coexisting state exists
at any energy.

We note that some of these parameters may be ill-defined for small clusters. Nonetheless
the values of $R_1$ and $R_2$ may be calculated using the capillarity approximation i.e.
using bulk values (if available) for the material parameters. For lead \cite{BenDavid95}
we find that $R_1 = 0.2$ nm (or approximately 1 atom) and $R_2 = 0.6$ nm (or approximately 100 atoms).
This suggests that coexistence ought to be stable down to quite small cluster sizes, although it is
well-known that the melting behaviour of sub-100 atom clusters is quite erratic due to electron-shell
effects \cite{Schmidt98} that are certainly not captured in the MCD model.

\section{Simulated Caloric Curves}

\begin{figure}
 \resizebox{\columnwidth}{!}{\includegraphics{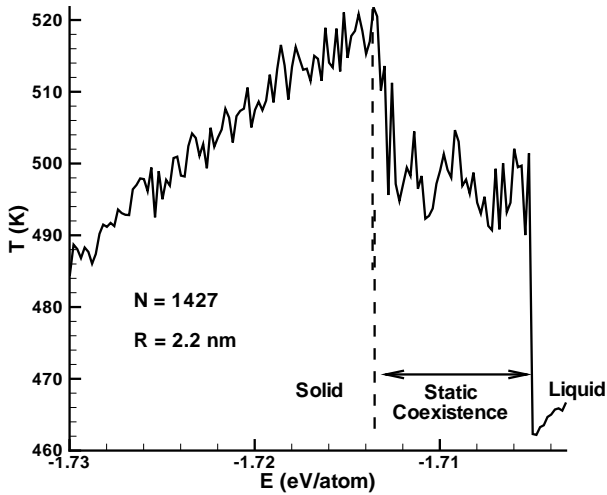}}  (a)
 \resizebox{\columnwidth}{!}{\includegraphics{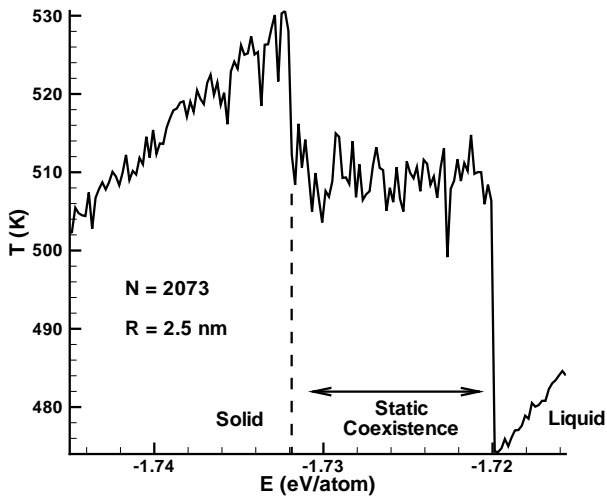}}  (b)
\caption{\label{caloric-MD} Caloric curves for
(a) a 1427-atom Pb cluster and (b) a 2073-atom cluster. The 1427 and 2073-atom
clusters exhibit coexistence for a range of energies.}
\end{figure}

Using molecular dynamics simulations of Pb clusters we have found that coexistence
becomes unstable at approximately 1000-atoms. We simulated the caloric curves for three
clusters: a 2073-atom icosahedron, a 1427-atom  icosahedron and a 931-atom icosahedron.
Here we use surface-reconstructed icosahedra which are thought to be stable in Pb
clusters \cite{Hendy01,Hendy02} at these sizes using a glue potential \cite{LOE92}. At each energy
the cluster was equilibrated for 0.6 ns, then the kinetic energy was averaged over a further 0.6 ns.
The energy increment used was 0.2 meV/atom with energies adjusted between constant energy
simulations by a uniform scaling of the kinetic energy. Figure~\ref{caloric-MD} shows the
caloric curves for the 2073 and 1427-atom clusters moving from the solid state (on left) to the
liquid state (on right). Both clusters exhibit a coexisting solid-liquid state separating
the solid and liquid phases. The curves in figure~\ref{caloric-MD} resemble those constructed
in Ref~\cite{Wales95} with their layer-by-layer model of cluster melting.

\begin{figure}
\resizebox{\columnwidth}{!}{\includegraphics{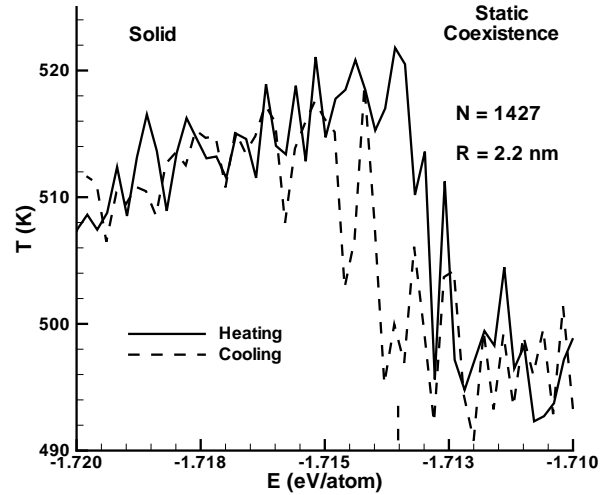}}
\caption{\label{meltfreeze} Caloric curves for a 1427-atom cluster produced first by heating
(solid line), then by cooling (dashed line).}
\end{figure}

Note the change in temperature at the transition from fully solid to coexistence and in the
transition from coexistence to fully liquid. In MCD theory, while the transition from coexistence
to liquid is first-order, the transition from solid to coexistence is continuous. To check whether
the 1427 and 2057-atom clusters are superheated when they undergo the transition from solid to
the coexisting state, we cooled a 1427-atom cluster from the coexistence region down until the
cluster froze. Figure~\ref{meltfreeze} shows the reverse caloric curve (constructed using
equilibration times of 4 ns with an energy increment of -0.2 meV/atom). The freezing transition
takes place to well within 5 meV/atom of the transition from solid to coexistence, suggesting that
the simulated caloric curve is close to the equilibrium caloric curve and that the transition at
-1.715 eV/atom is a first-order transition which exhibits little rounding due to finite-size effects.
The origin of this discontinuity at this transition may correspond to the creation of a ``minimum"
volume of liquid and as such the discontinuity is likely to disappear as $R \rightarrow \infty$.

\begin{figure}
\resizebox{\columnwidth}{!}{\includegraphics{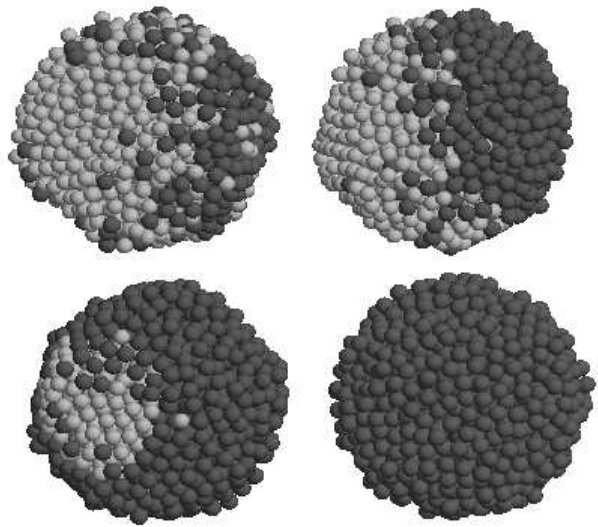}}
\caption{\label{mixed} A sequence of snapshots showing the coexisting solid-liquid state as the
energy is increased from $E= -1.712$ eV/atom (top left) to $E=-1.706$ eV/atom (bottom right). Atoms
are classified by their mobilities: the dark grey atoms are liquid (high mobility) while the light
grey atoms are solid (low mobility).}
\end{figure}

To characterise the coexisting state, we follow Cleveland et al \cite{Cleveland94} in using the
bimodality of the distribution of diffusion coefficients to distinguish solid and liquid atoms.
A sequence of snapshots at different total energies of the 1427-atom cluster is shown in
figure~\ref{mixed}. The total energy of the atoms in each snapshot decreases from top left to top
right to bottom left to bottom right. Each atom in the snapshot has been coloured either white or
grey depending on the whether the diffusion coefficient of that atom (calculated over a carefully
chosen timescale) is low mobility (solid) or high mobility (liquid). The snapshots reveal that the
molten lead does not fully wet the solid. They also reveal that the liquid fraction of the coexisting
cluster increases as the total energy increases. Note that the melt does not wet the solid as assumed
in equation~\ref{energy} whereas in bulk lead, the melt wets the solid (since in bulk lead
$\Delta \gamma > 0$). This is a clear indication that the capillarity approximation has broken down
at these cluster sizes.

\begin{figure}
\resizebox{\columnwidth}{!}{\includegraphics{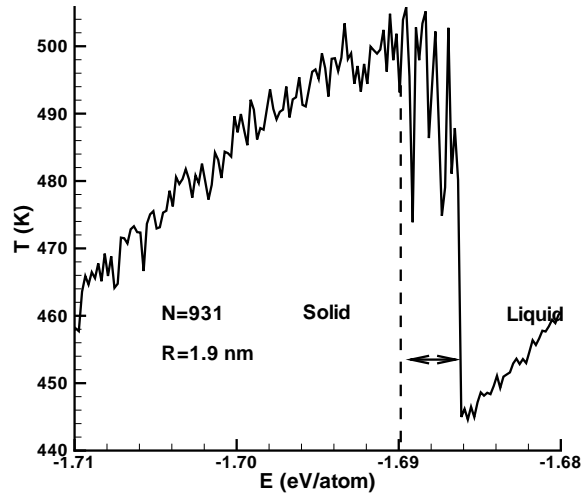}}
\caption{\label{caloric-931} Caloric curves for a 931-atom Pb cluster. The 931-atom cluster does
not exhibit a stable coexisting phase. The large fluctuations near the melting point signal
the appearance of precritical liquid nuclei. The true melting point energy probably lies closer to
the dotted line.}
\end{figure}

Figure~\ref{caloric-931} shows the caloric curve for a 931-atom cluster constructed in the same way
as those for the larger clusters in figure~\ref{caloric-MD}. Note that the curve does not show a
stable solid-liquid coexisting state but that large fluctuations in temperature occur near the
melting transition in the 931-atom caloric curve. Using the same method for identifying solid and
liquid regions as was used in figure~\ref{mixed}, one can verify that the fluctuations in temperature
are correlated with the appearance of a liquid region in the otherwise solid cluster. This is
due to the dynamic coexistence between the solid state and a coexisting state.

\begin{figure}
\resizebox{\columnwidth}{!}{\includegraphics{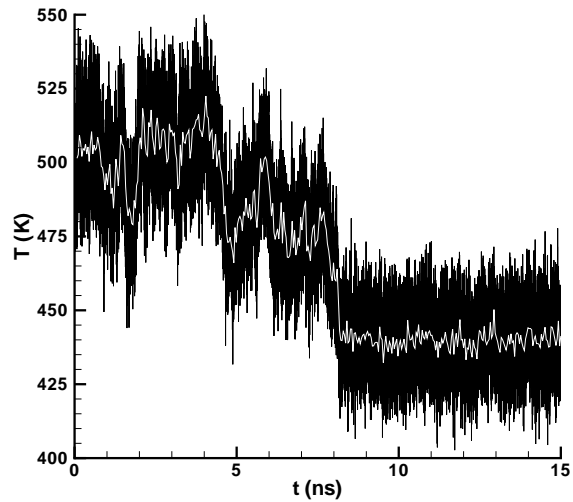}}
\caption{\label{trace-931} Time evolution of the kinetic temperature in a 931-atom
cluster at an energy $E=-1.688$ eV/atom. The white line shows a time averaged
trace of the temperature. The cluster starts as solid (0-5 ns), becomes coexisting
(5-8 ns) and finally melts fully (after 8 ns).}
\end{figure}

Which state is globally stable? We have resolved this with a very long duration run near the melting
transition. Figure~\ref{trace-931} shows a history of the temperature (as determined using the method above)
for a 931-atom cluster at a total energy of $E=-1.688$ eV/atom over approximately 15 ns. The final state of the cluster is liquid (achieved after approximately
8 ns), which suggests that the fluctuations (between 0 and 8 ns in figure~\ref{trace-931} and
between $-1.690$ and $-1.686$ eV/atom in figure~\ref{caloric-931}) are non-equilibrium precursors
to melting rather than true dynamic coexistence. MCD theory certainly suggests that at
sizes $R < R_2$ the coexisting state is metastable with respect to the liquid rather than the solid.
However, the caloric curve in figure~\ref{caloric-931} suggests that the coexisting state is
metastable with respect to both the solid and liquid states, as the cluster appears to spend more
time in the solid state. Nonetheless, we conclude that there is no stable coexisting state for the
931-atom cluster at any energy. In Ref~\cite{Hendy03}, we constructed a less detailed caloric curve
for 1130-atom lead clusters by coalescence. This curve also showed no evidence for stable coexistence.

\section{Discussion}

The molecular dynamics simulations and MCD theory in the capillary approximation do
not agree on the size at which coexistence becomes metastable in Pb clusters. However,
the snapshots in figure~\ref{mixed} strongly suggest that the capillarity approximation
has broken down at these cluster sizes (at least when it comes to reproducing the behaviour
of the simulated clusters). This should not be surprising as atomic clusters are interesting
precisely because their properties depend strongly on size. However, if we are to take MCD theory
beyond the capillarity approximation, we would require a detailed understanding
of how surface energies, and other quantities such as latent heat, depend on surface
curvature and/or size. It is not simply a matter of inserting more accurate material properties
into (\ref{R2}).

It is also important to recognise that the spherically symmetric geometry assumed in equation
(\ref{energy}) is inappropriate for situations involving partial wetting where
$\Delta \gamma < 0$. In Ref~\cite{Cleveland94}, a critical droplet model was developed for
the partial wetting case, although it was applied in the canonical ensemble. Evidently a
generalisation of MCD theory to the partially wetting case would also be desirable. Similarly
it would be interesting to determine by simulation and experiment the size at which
$\Delta \gamma = 0$ in Pb clusters i.e. the size below which complete wetting gives way to partial
wetting. Partial and complete wetting has been observed recently in isolated metal alloy clusters
using transmission electron microscopy \cite{Lee04}.

Finally, in the simulations  the transition from solid to solid-liquid coexistence
was discontinuous. In MCD theory this transition is continous as the energy of the coexisting
cluster, as given by equation (\ref{energy}), approaches the energy of an entirely solid cluster
as $\eta \rightarrow 1$. This limit is an artificial feature of equation (\ref{energy}); if this
constraint were to be relaxed so that the energy of the coexisting cluster
approached some other value as $\eta \rightarrow 1$ (for instance,
$e_s + 3 (\gamma_{lv}+\gamma_{sl})/ \rho R)$ then the transition from solid to solid-liquid would be
discontinuous. Provided the energy of the solid cluster and that of the limiting solid-liquid cluster
differed only by surface energy, this discontinuity would disappear as $R \rightarrow \infty$.

\section{Conclusion}

In summary, MCD theory predicts that the static coexistence of solid and liquid phases can
only occur in clusters above a certain size given by equation (\ref{R2}). Below this critical
size coexistence is metastable or unstable. Molecular dynamics simulations of Pb clusters
using a glue potential appear to corroborate this scenario, although the transition occurs at a
larger size, $N_2$,  (where $1100 < N_2 < 1400$) than that predicted by MCD theory using bulk material
parameters in the capillarity approximation. Indeed this discrepancy between theory and simulation
is probably due to the breakdown of this approximation at small cluster sizes. We note that Pb has
an unusually large value of the prefactor $x = (\gamma_{sv}-\gamma_{lv})/ \rho L$ that appears in
equation (\ref{R2}) due to a relatively small latent heat of fusion. Other metals with small latent
heats (which are correlated with low melting points), such as Hg or the alkali metals, will also have
large values of $x$. It may be that in such metals coexistence is forbidden at quite large sizes.
In most metals, however, with relatively large latent heats of fusion, we might expect static
coexistence to remain stable down to rather small sizes.

\begin{acknowledgments}
The author would like to thank Dr Jonathan Doye for some useful comments regarding this
work. This work was funded by the MacDiarmid Institute for Advanced Materials and
Nanotechnology.
\end{acknowledgments}



\begin{thebibliography}{27}
\expandafter\ifx\csname natexlab\endcsname\relax\def\natexlab#1{#1}\fi
\expandafter\ifx\csname bibnamefont\endcsname\relax
  \def\bibnamefont#1{#1}\fi
\expandafter\ifx\csname bibfnamefont\endcsname\relax
  \def\bibfnamefont#1{#1}\fi
\expandafter\ifx\csname namefont\endcsname\relax
  \def\citenamefont#1{#1}\fi
\expandafter\ifx\csname url\endcsname\relax
  \def\url#1{\texttt{#1}}\fi
\expandafter\ifx\csname urlprefix\endcsname\relax\def\urlprefix{URL }\fi
\providecommand{\bibinfo}[2]{#2}
\providecommand{\eprint}[2][]{\url{#2}}

\bibitem[{\citenamefont{Pochon et al}(2004)}]{Pochon04}
  \bibinfo{author}{\bibfnamefont{S.} \bibnamefont{Pochon}},
  \bibinfo{author}{\bibfnamefont{K.~F.} \bibnamefont{MacDonald}},
  \bibinfo{author}{\bibfnamefont{R.~J.} \bibnamefont{Knize}} \bibnamefont{and}
  \bibinfo{author}{\bibfnamefont{N.~I.} \bibnamefont{Zheludev}},
  \bibinfo{journal}{Phys. Rev. Lett.} \textbf{\bibinfo{volume}{92}},
  \bibinfo{pages}{145702} (\bibinfo{year}{2004}).

\bibitem[{\citenamefont{Lee and Mori}(2004)}]{Lee04}
  \bibinfo{author}{\bibfnamefont{J.-G.} \bibnamefont{Lee}} \bibnamefont{and}
  \bibinfo{author}{\bibfnamefont{H.} \bibnamefont{Mori}},
  \bibinfo{journal}{Phys. Rev. B.} \textbf{\bibinfo{volume}{70}},
  \bibinfo{pages}{144105} (\bibinfo{year}{2004}).

\bibitem[{\citenamefont{Honeycutt and Andersen}(1987)}]{Honeycutt87}
\bibinfo{author}{\bibfnamefont{J.~D.} \bibnamefont{Honeycutt}}
\bibnamefont{and}
\bibinfo{author}{\bibfnamefont{H.~C.} \bibnamefont{Andersen}},
\bibinfo{journal}{J. Phys. Chem.} \textbf{\bibinfo{volume}{91}},
  \bibinfo{pages}{4950} (\bibinfo{year}{1987}).

\bibitem[{\citenamefont{Reiss et~al.}(1988)}]{Reiss88}
\bibinfo{author}{\bibfnamefont{H.} \bibnamefont{Reiss}}
\bibinfo{author}{\bibfnamefont{P.} \bibnamefont{Mirabel}},
\bibnamefont{and}
\bibinfo{author}{\bibfnamefont{R.~L.} \bibnamefont{Whetten}},
\bibinfo{journal}{J. Phys. Chem.} \textbf{\bibinfo{volume}{92}},
  \bibinfo{pages}{7241-7246} (\bibinfo{year}{1988}).

\bibitem[{\citenamefont{Wales and Berry}(1994)}]{Wales94}
  \bibinfo{author}{\bibfnamefont{D.~J.} \bibnamefont{Wales}} \bibnamefont{and}
  \bibinfo{author}{\bibfnamefont{R.~S.} \bibnamefont{Berry}},
  \bibinfo{journal}{Phys. Rev. Lett.} \textbf{\bibinfo{volume}{73}},
  \bibinfo{pages}{2875} (\bibinfo{year}{1994}).

\bibitem[{\citenamefont{Nielsen et~al.}(1994)}]{Nielsen94}
\bibinfo{author}{\bibfnamefont{O.~H.} \bibnamefont{Nielsen}},
\bibinfo{author}{\bibfnamefont{J.~P.} \bibnamefont{Sethna}},
\bibinfo{author}{\bibfnamefont{P.} \bibnamefont{Stoltze}},
\bibinfo{author}{\bibfnamefont{K.~W.} \bibnamefont{Jacobsen}},
\bibnamefont{and}
  \bibinfo{author}{\bibfnamefont{J.~K.} \bibnamefont{Norskov}},
  \bibinfo{journal}{Europhys. Lett.} \textbf{\bibinfo{volume}{26}},
  \bibinfo{pages}{51-56} (\bibinfo{year}{1994}).

\bibitem[{\citenamefont{Cleveland et~al.}(1994)}]{Cleveland94}
\bibinfo{author}{\bibfnamefont{C.~L.} \bibnamefont{Cleveland}},
\bibinfo{author}{\bibfnamefont{U.} \bibnamefont{Landman}},
\bibnamefont{and}
\bibinfo{author}{\bibfnamefont{W.~D.} \bibnamefont{Luedtke}},
  \bibinfo{journal}{J. Phys. Chem.} \textbf{\bibinfo{volume}{98}},
  \bibinfo{pages}{6272-6279} (\bibinfo{year}{1994}).

\bibitem[{\citenamefont{Lynden-Bell and Wales}(1994)}]{Lynden-Bell}
  \bibinfo{author}{\bibfnamefont{R.~M.} \bibnamefont{Lynden-Bell}}
  \bibnamefont{and} \bibinfo{author}{\bibfnamefont{D.~J.} \bibnamefont{Wales}},
  \bibinfo{journal}{J. Chem. Phys.} \textbf{\bibinfo{volume}{101}},
  \bibinfo{pages}{1460} (\bibinfo{year}{1994}).

\bibitem[{\citenamefont{Schmidt et~al.}(2001)}]{Schmidt01}
\bibinfo{author}{\bibfnamefont{M.} \bibnamefont{Schmidt}} \bibnamefont{{\em et~al.}},
\bibinfo{journal}{Phys. Rev. Lett.} \textbf{\bibinfo{volume}{86}},
  \bibinfo{pages}{1191-1194} (\bibinfo{year}{2001}).

\bibitem[{\citenamefont{Cleveland et~al.}(1999)\citenamefont{Cleveland,
  Luedtke and Landman}}]{Cleveland99}
\bibinfo{author}{\bibfnamefont{C.~L.} \bibnamefont{Cleveland}},
\bibinfo{author}{\bibfnamefont{W.~D.}~\bibnamefont{Luedtke}},
\bibnamefont{and}
\bibinfo{author}{\bibfnamefont{U.}~\bibnamefont{Landman}},
\bibinfo{journal}{Phys. Rev. B} \textbf{\bibinfo{volume}{60}},
\bibinfo{pages}{5065-5077} (\bibinfo{year}{1999}).

\bibitem[{\citenamefont{Gross}(2001)}]{Gross01}
\bibinfo{author}{\bibfnamefont{D.H.E.}~\bibnamefont{Gross}},
  \emph{\bibinfo{title}{Microcanonical thermodynamics: phase transitions in finite systems}}
  \bibinfo{journal}{Lecture notes in Physics} \textbf{\bibinfo{volume}{66}},
  \bibinfo{publisher}{World Scientific},(\bibinfo{year}{2001}).

\bibitem[{\citenamefont{Ben David et~al.}(1995)}]{BenDavid95}
\bibinfo{author}{\bibfnamefont{T.} \bibnamefont{Ben David}}  \bibnamefont{{\em et~al.}},
  \bibinfo{journal}{Phil. Mag. A} \textbf{\bibinfo{volume}{71}},
  \bibinfo{pages}{1135-1143} (\bibinfo{year}{1995}).

\bibitem[{\citenamefont{Schmidt et~al.}(1998)}]{Schmidt98}
\bibinfo{author}{\bibfnamefont{M.} \bibnamefont{Schmidt}}
 \bibnamefont{{\em et~al.}},
\bibinfo{journal}{Nature} \textbf{\bibinfo{volume}{393}},
  \bibinfo{pages}{238-240} (\bibinfo{year}{1998}).

\bibitem[{\citenamefont{Hendy and Hall}(2001)}]{Hendy01}
\bibinfo{author}{\bibfnamefont{S.~C.}~\bibnamefont{Hendy}} \bibnamefont{and}
  \bibinfo{author}{\bibfnamefont{B.~D.}~\bibnamefont{Hall}},
  \bibinfo{journal}{Physical Review B} \textbf{\bibinfo{volume}{64}},
  \bibinfo{pages}{085425} (\bibinfo{year}{2001}).

\bibitem[{\citenamefont{Hendy and Doye}(2002)}]{Hendy02}
 \bibinfo{author}{\bibfnamefont{S.~C.} \bibnamefont{Hendy}} \bibnamefont{and}
\bibinfo{author}{\bibfnamefont{J.~P.~K.} \bibnamefont{Doye}}  ,
  \bibinfo{journal}{Physical Review B}  \textbf{\bibinfo{volume}{66}},
  \bibinfo{pages}{235402} (\bibinfo{year}{2002}).

\bibitem[{\citenamefont{Lim et~al.}(1992)\citenamefont{Lim, Ong, and
  Ercolessi}}]{LOE92}
\bibinfo{author}{\bibfnamefont{H.~S.} \bibnamefont{Lim}},
  \bibinfo{author}{\bibfnamefont{C.~K.} \bibnamefont{Ong}}, \bibnamefont{and}
  \bibinfo{author}{\bibfnamefont{F.}~\bibnamefont{Ercolessi}},
  \bibinfo{journal}{Surface Science} \textbf{\bibinfo{volume}{269/270}},
  \bibinfo{pages}{1109} (\bibinfo{year}{1992}).

\bibitem[{\citenamefont{Wales and Doye}(1995)}]{Wales95}
\bibinfo{author}{\bibfnamefont{D.~J.} \bibnamefont{Wales}} \bibnamefont{and}
  \bibinfo{author}{\bibfnamefont{J.~P.~K.} \bibnamefont{Doye}},
  \bibinfo{journal}{J. Chem. Phys.} \textbf{\bibinfo{volume}{103}},
  \bibinfo{pages}{3061-3070} (\bibinfo{year}{1995}).

\bibitem[{\citenamefont{Hendy et~al.}(2003)}]{Hendy03}
\bibinfo{author}{\bibfnamefont{S.}~\bibnamefont{Hendy}},
\bibinfo{author}{\bibfnamefont{S.~A.}~\bibnamefont{Brown}} \bibnamefont{and}
  \bibinfo{author}{\bibfnamefont{M.}~\bibnamefont{Hyslop}},
  \bibinfo{journal}{Physical Review B} \textbf{\bibinfo{volume}{68}},
  \bibinfo{pages}{241403(R)} (\bibinfo{year}{2003}).

\end{thebibliography}
\end{document}